%
%

\magnification 1200
\baselineskip=18pt


\centerline{\bf SUBLIMATED DECOUPLING OF THE VORTEX LATTICE}
\bigskip
\centerline{\bf IN EXTREMELY TYPE-II LAYERED SUPERCONDUCTORS}
\vskip 50pt


\centerline{J. P. Rodriguez}
\medskip
\centerline{\it Dept. of Physics and Astronomy,
California State University, Los Angeles, CA 90032.}
\vskip 30pt
\centerline  {\bf  Abstract}
\vskip 8pt\noindent
The question of whether layer decoupling and vortex-lattice
melting occur separately or not 
in the mixed phase of pristine 
layered superconductors in the  extreme type-II
limit is studied through a
partial duality analysis of the layered $XY$ model with
uniform frustration.  We find that both transitions occur
simultaneously if the normal/superconducting transition of
the vortex lattice in an isolated layer is first order
and  if a sufficient degree of layer anisotropy exists.
We    also find       that a crossover to a highly entangled
vortex lattice phase with relatively low phase rigidity across layers does
not occur in practice
under any circumstances
at  temperatures below the  two-dimensional vortex-lattice melting transition.

\bigskip
\noindent
PACS Indices:  74.60.Ge, 74.60.-w,  74.60.Ec, 74.25.Dw

\vfill\eject







\centerline
{\bf I.  Introduction}
\bigskip

It is now well established experimentally that the Abrikosov vortex lattice
state in clean high-temperature superconductors undergoes a first-order
melting transition into a liquid phase.$^{1}$
High-temperature superconductors are layered and extremely type-II.$^2$
The former vortex liquid phase in the most anisotropic materials like
BSCCO is best described by a liquid of 
planar vortices inside of decoupled layers.$^3$
A longstanding question 
 is whether melting and layer decoupling occur simultaneously as a
sublimation transition, or whether a separate
decoupling transition follows the melting transition.
Some experimental studies on the highly anisotropic
BSCCO material show evidence for sublimation,$^{4,5}$
while most experimental studies of the less anisotropic
YBCO material point to separate melting and decoupling
transitions.$^{6}$

The  experimental situation outlined above suggests that 
the degree of anisotropy
is what in fact determines whether or not the vortex lattice
in a layered superconductor sublimates.
We shall study this proposal theoretically by 
analyzing the layered $XY$ model with uniform frustration,
which provides a qualitatively correct description 
of the thermodynamics deep inside of the mixed phase
in extremely type-II layered superconductors.$^7$
After performing  a partial duality transformation on the $XY$ model
that is particularly well suited to the weak-coupling limit,$^8$
we find that there can exist as many as three different decoupling 
transitions 
 at temperatures $T_D < T_m < T_{\times}$, respectively.
(We use the term `transition' here loosely to describe both
genuine phase transitions and cross-overs.)
The phase correlation length across  layers is equal to the inter-layer
spacing along  the 
dimensional crossover line$^{9,10}$  at $T = T_{\times}$  
that separates two-dimensional (2D)
from three-dimensional (3D) vortex-liquid behavior.$^6$
The  phase correlation length across layers then
either  diverges or jumps to infinity along the melting line, $T = T_m$,
which separates the superconducting and normal phases.
Last, the crossover  line $T = T_D$ that lies inside of
the ordered phase is defined by the point at which the
Josephson coupling energy reaches about half of its zero-temperature value.
The macroscopic phase rigidity across layers becomes small
in comparison to its zero-temperature value at
temperatures $T >   T_D$ because of the
entanglement of fluxlines between adjacent layers.$^{8,11}$
All three decoupling transitions occur separately in the
continuum regime at low perpendicular vortex density,
but $T_D$ crosses below the 
2D melting temperature at
only exponentially weak inter-layer coupling.
At a moderate concentration of vortices, on the other hand,
we find that the three decoupling transitions collapse
onto a single sublimation line for weak enough Josephson
coupling.  This is due 
to the 
first-order nature of the melting of the 2D vortex lattice in such case.
These results are compared with previous theoretical
calculations based on the elastic medium description of the
vortex lattice$^{2, 3, 11, 12}$ 
and with direct Monte Carlo simulation results 
of the $XY$ model itself.$^{7}$

\bigskip
\centerline
{\bf II.  Duality Theory}
\bigskip

The layered $XY$ model with uniform frustration is the minimum
theoretical description of  vortex matter in 
extremely type-II layered
superconductors.  Both fluctuations of the magnetic induction and
of the magnitude of the superconducting order parameter are neglected
within this approximation.
The model hence  is valid deep inside
the interior of the mixed phase.  The thermodynamics of the
3D $XY$ model
with anisotropy and uniform frustration
is determined by the superfluid kinetic energy
$E_{XY}^{(3)} = -\sum_{r,\mu} J_{\mu} {\rm cos}
[\Delta_{\mu}\phi  - A_{\mu}]|_r$,
which is a  functional of the superconducting
phase $\phi(r)$ over the cubic lattice.  
Here, $J_x = J = J_y$ and $J_z = J/\gamma^{\prime 2}$
are the local phase rigidities, with anisotropy
parameter $\gamma^{\prime} > 1$.
The   vector potential
$A_{\mu} = (0, 2\pi f x/a, 0)$
represents the magnetic induction
oriented perpendicular to the layers,
$B_{\perp} = \Phi_0 f / a^2$.
Here $a$ denotes the square lattice constant, which is of order
the zero-temperature coherence length, $\Phi_0$ denotes
the flux quantum, and $f$ denotes the concentration of vortices per site.    
The component of the magnetic induction parallel to the
layers in taken to be null throughout.

We shall now analyze the above layered system in the selective
high-temperature limit, $k_B T\gg J_z$.
Following ref. 8, the corresponding  high-temperature
expansion can be achieved through  a {\it partial} duality transformation
of the layered $XY$ model
along the $z$ axis  perpendicular to the layers.
This leads to a useful layered Coulomb gas (CG) ensemble in terms
of loops of Josephson vortices in between layers (fluxons).$^{13}$
In particular, suppose that $l$ denotes the layer index,
that $\vec r$ represents the x-y coordinates, and that
$r = (\vec r, l)$.
Phase correlations across $N$ layers are then described 
by the phase auto-correlation function probed at sites
set by an  integer field 
$p(r) = \delta_{\vec r, 0} (\delta_{l, 1} - \delta_{l, N})$.
These can be computed
from the quotient
$$\Bigl\langle {\rm exp} \Bigl[i\sum_r p(r) \phi(r)\Bigr]\Bigr\rangle =
Z_{\rm CG}[p]/Z_{\rm CG}[0]\eqno (1)$$
of partition functions for
a    layered CG ensemble that
describes the nature of the Josephson coupling:$^8$
$$Z_{\rm CG}[p] = \sum_{\{n_{z}(r)\}} y_0^{N[n_z]}\cdot
\Pi_{l} C     [q_l]\cdot
e^{-i\sum_r n_z A_z},  \eqno (2)$$
where $n_z (\vec r, l)$ is an integer field
on links between adjacent layers $l$ and $l+1$
located  at 2D points $\vec r$.
The ensemble is weighted
by a product
of phase auto-correlation functions
$$C   [q_l] =
\Bigl\langle {\rm exp} \Bigl[i \sum_{\vec r}
q_l (\vec r) \phi (\vec r, l)\Bigr]\Bigr\rangle_{J_z = 0}
\eqno (3)$$
for isolated layers $l$
probed at the dual   charge  that accumulates onto
that layer:
$$q_l (\vec r) = p(\vec r, l) +  n_z (\vec r, l-1) - n_z (\vec r, l).
\eqno (4)$$
It is also weighted
by a bare fugacity
$y_0$   that is
raised to the power
$N [n_z]$
equal to the total
 number of dual charges, $n_z = \pm 1$.
The fugacity is
given by 
$y_0 = J_z / 2 k_B T$ in the selective high-temperature regime,
$J_z \ll k_B T$,
reached at large model  anisotropy.
Also, the average number of $n_z$ charges per link is equal to$^8$
$2 y_0 (\langle {\rm cos}\, \phi_{l,l+1}\rangle - y_0)$,
which is less than $J_z/k_B T$.
This implies  that   the layered  CG ensemble (2) is dilute
in such case, because $y_0\ll 1$. 
The former is required by the approximate nature of Eq. (2),
which neglects multiple occuppancy of the dual charges, $n_z$, on
a given link.
Last, the thermodynamics of the layered $XY$ model is encoded
by its partition function, which is given by the following
product:
$$Z_{XY}^{(3)}[0]  = [I_0 (J_z/k_B T)]^{{\cal N}^{\prime}}\cdot
Z_{\rm CG}[0]\cdot \Pi_{l} Z_{XY}^{(2)}[0].\eqno (5)$$
Here, $I_0 (x)$ is a modified Bessel function, and 
$Z_{XY}^{(2)}[0]$ is the partition function of an isolated layer.
Also, ${\cal N}^{\prime}$ denotes the total  number of links between
adjacent layers.

Interlayer correlations of the layered $XY$ are easily determined
using the CG ensemble (2) when
the phase correlations within an isolated layer are short range.$^{8}$  
Let us introduce  the notation
$\phi_{l,l^{\prime}} (\vec r) = \phi (\vec r, l^{\prime}) -
 \phi (\vec r, l)$
and take $A_z = 0$ due to the  null magnetic 
field parallel to the layers. 
A  useful (in)equality for the autocorrelator between
any number of layers, $n+1$,
can be computed to lowest order in the fugacity, $y_0$. 
It reads$^{8}$
$$\langle e^{i\phi_{l,l+n}}\rangle \leq
\Biggl[{\cal C}_0 \int {d^2 q\over{(2\pi)^2}}
\Biggl({{\cal C}_q\over{{\cal C}_0}}\Biggr)^{n+1}\Biggr]
 (y_0 {\cal C}_0/a^2)^n , \eqno (6)$$
where
$${\cal C}_q =
\int d^2r |C (\vec r)|
 e^{i\vec q\cdot\vec r}\eqno (7)$$
is the Fourier transform of 
the magnitude 
of the phase auto-correlation function
(3)
for an isolated layer probed at two points,
$\vec r_1$ and $\vec r_2$:
$$C   (1, 2) = |C (\vec r_{12})|
e^{-i\int_1^2 \vec A^{\prime}(\vec r)\cdot d\vec r},\eqno (8)$$
where $\vec  A^{\prime}$
is a suitably  gauge-transformed vector potential (see below).
Its magnitude depends only on the separation 
$\vec r_{12} = \vec r_1 - \vec r_2$ between the probes,
and it decays exponentially at separations beyond a characteristic
correlation length $\xi_{2D}$ due to the 
phase-incoherent state that is presently assumed.
The layered CG ensemble (2) is therefore in a   confining phase.$^{14}$ 
The prefactor in brackets above in Eq. (6) typically decays polynomially
with the  separation $n$ between layers.
Also, Eq. (6) is an equality 
for $n = 1$,$^{15}$  as well as for
pure gauges such that
$\vec A^{\prime} = \vec\nabla \phi_0$ (see below). 
To conclude, the autocorrelator
$\langle e^{i\phi_{l,l+n}}\rangle$
across layers
decays at least exponentially with the separation
 $n$ in the weak-coupling limit,
$y_0\rightarrow 0$, of the disordered phase.

The layered CG ensemble (2) can also be used 
to determine interlayer correlations
in the ordered phase.  Consider again an isolated layer,
and suppose that
general phase auto-correlation
functions (3)  are quasi-long range:
$$C   [q] = g_0^{n_+}\cdot {\rm exp}\Bigl[  \eta_{2D}
\sum_{(1,2)} q(1){\rm ln} (r_{12} / r_0)\, q(2)\Bigr] \cdot
{\rm exp} \Bigl(i\sum q \cdot \phi_0 \Bigr),\eqno (9)$$
where $g_0$ 
is equal to the phase rigidity of an isolated layer in units of $J$,
where $n_+$ is equal to half the number probes,
where $r_0$ is the natural ultraviolet scale
of order the inter-vortex spacing,
$a_{\rm vx} = a/f^{1/2}$, 
and where $\phi_0 (\vec r)$ should resemble the unique
zero-temperature configuration 
(independent of the layer index, $l$).
The system of dual ($n_z$) charges in the layered CG ensemble (2) is 
then in a   plasma phase
at low temperatures $\eta_{2D} < 2$.$^{8, 13}$
In such case,
the macroscopic phase rigidity across layers
is approximately given by$^8$
$$\rho_s^{\perp} / J_{z} \cong
\langle {\rm cos}\, \phi_{l,l+1}\rangle - y_0.
\eqno (10)$$
Furtheremore, in this case
an appropriate Hubbard-Stratonovich transformation
of the CG partition function (2) in the absence 
of a source ($p = 0$) 
reveals that
it is equal
to the corresponding one
$Z_{\rm LD}[0] = \int {\cal D} \theta\,  e^{-E_{\rm LD}/k_B T}$
for a   renormalized Lawrence-Doniach (LD) model
up to a factor that is independent of the Josephson coupling, $J_z$.
The corresponding energy functional is given by$^8$
$$\eqalignno{
E_{\rm LD} = &
\bar J\int d^2 r \Biggl[
\sum_{l}
{1\over 2}(\vec\nabla\theta_l)^2
-\Lambda_0^{-2}
\sum_{l}{\rm cos} (\theta_{l+1} - \theta_l)\Biggr],
&(11)\cr}$$
where $\bar J = k_B T / 2\pi \eta_{2D}$
is the  macroscopic  phase rigidity of an isolated layer,$^{18}$
and where $\Lambda_0 = \gamma^{\prime} a$ is the Josephson
penetration length.  
The above continuum
description (11) is understood to have an  ultraviolet
cut off  of order the inter-vortex spacing,
$r_0$. 
A standard analysis of the product of partition functions 
 (5) then yields that 
 the strength of the local Josephson coupling is 
given by
$$\langle {\rm cos}\, \phi_{l,l+1}\rangle  =
 y_0
 +  g_0 \langle {\rm cos}\, \theta_{l,l+1}\rangle,
 \eqno (12)$$
where $\theta_{l,l+1}  = \theta_{l+1} - \theta_l$.

To compute $\langle {\rm cos}\, \theta_{l,l+1}\rangle$ 
in the weak-coupling limit,
it is sufficient to consider only layers $l$ and $l+1$ in isolation
from the rest of the system. 
At low temperature $\eta_{2D}\ll 1$, the harmonic approximation
for the Josephson coupling term   in Eq. (11) is valid:
${\rm cos}\, \theta_{l,l+1} \cong 1 - {1\over 2} \theta_{l,l+1}^2$.
The resulting gaussian integration then yields
$\langle{\rm cos}\, \theta_{l,l+1}\rangle = 
e^{-\langle \theta_{l,l+1}^2\rangle / 2}$,
with 
$\langle \theta_{l,l+1}^2\rangle =
\eta_{2D} \, {\rm ln} (\Lambda_J^2/ r_0^2)$.  
Here $\Lambda_{J}$ is of order the Josephson pentration
length, $\Lambda_0 = \gamma^{\prime} a$.  
Subsitution into Eq. (12)
then produces the result$^8$
$$\langle {\rm cos}\, \phi_{l,l+1}\rangle  =
 y_0
 +  g_0  (r_0/\Lambda_J)^{\eta_{2D}}
 \eqno (13)$$
for  the strength of the local Josephson coupling 
at low temperature $\eta_{2D}\ll 1$.
The latter agrees with the result produced by analyzing a
fermion analogy for the LD model (11),
as well as with an estimate by Glazman and Koshelev
for the zero-field case ($r_0\sim a$).$^{10}$
Subsitution of this result into  Eq.  (10) therefore yields
the formula
$$\rho_s^{\perp}/J_z = g_0 (r_0/\Lambda_J)^{\eta_{2D}}\eqno (14)$$
for the macroscopic phase rigidity across layers
in this regime.$^8$  To conclude, 
macroscopic 
phase coherence    exists across
layers in the ordered phase (9).

\bigskip
\centerline
{\bf III.  Continuum Limit}
\bigskip

We shall now  review the phase diagram
that results from employing the above duality analysis for the layered
$XY$ model in the continuum limit,$^8$ 
 $a\rightarrow 0$,
which coincides  with the
regime of small perpendicular flux density, $f\ll 1/36$.
In the absence of surface barriers,
Monte Carlo simulations$^{16}$
indicate that the vortex liquid phase of an isolated layer 
solidifies into a ``floating'' vortex lattice phase at
the 2D melting temperature, $k_B T_m^{(2D)} \cong  J/20$.  
A recent duality analysis
of such a single layer finds that the standard     2D melting scenario$^{17}$
takes place as long as rigid translations of the 2D vortex lattice
are prohibited by surface barriers.$^{19}$  
In particular, general 
phase auto-correlation
functions follow the form (9) in the vortex lattice
phase at $T < T_m^{(2D)}$,
with a 2D correlation exponent
that takes on an {\it extremely}
small value$^{19}$
$\eta_{2D} \cong (28\pi)^{-1}$ just below the 2D melting temperature,
$T_m^{(2D)}$.
Further, $\eta_{2D}$ decreases
linearly  to zero with decreasing temperature 
 in the 2D vortex lattice.
On the otherhand, the phase auto-correlation
function (8) decays exponentially with separation
 in the {\it hexatic} phase
that lies at temperatures just above $T_m^{(2D)}$.
The associated correlation length, $\xi_{2D}$,
diverges exponentially as temperature cools down to
$T_m^{(2D)}$.
The auto-correlation function retains, however,
the  trivial phase factor of the 2D vortex lattice:$^{19}$ 
$\int_1^2 \vec A^{\prime}\cdot d\vec r = \phi_0 (2) - \phi_0 (1)$.

We now illustrate that there exist as many as three
distinct decoupling temperatures:$^8$ $T_{\times} >  T_m > T_D$.
Consider the weak-coupling limit of the layered $XY$ model,
$\gamma^{\prime}\rightarrow\infty$.
Eq. (6) then becomes an equality in the hexatic phase of an isolated layer
due to the trivial phase factor in the phase auto-correlation
function (8).$^{19}$
The phase correlation length across  layers, $\xi_{\perp}$,
is therefore equal to the spacing $d$ between adjacent layers when
$$e^{-1} = y_0 \int d^2r |C (\vec r)| / a^2. \eqno (15)$$
This defines a dimensional  cross-over field,$^{3, 8-10}$
$$f \gamma_{\times}^{\prime 2} \sim 
g_0 (J/k_B T) (\xi_{2D}/a_{\rm vx})^2 \eqno (16)$$
in units of the naive  decoupling scale
$\Phi_0/\Lambda_0^2$,
that separates 2D from 3D vortex-liquid behavior.$^6$
It is traced  out in Fig. 1.
In these units, to be used hereafter, $f\gamma^{\prime 2}$
gives the perpendicular field. 
The system is best described by a decoupled stack of
2D vortex liquids at fields above $f \gamma_{\times}^{\prime 2}$.
On the ordered side at $T < T_m^{(2D)}$,
Eq.   (14) for $\rho_s^{\perp}$
 implies   that long-range order across layers exists:
$\xi_{\perp} = \infty$.
And since $g_0 J$ is equal to the phase rigidity of an isolated layer,
Eq. (14) also implies that 3D scaling is violated
at weak-coupling,
$(r_0/\Lambda_J)^{\eta_{2D}} \ll 1$,
in which case 
the phase rigidity across layers, 
$\rho_s^{\perp}$,
is small in comparison to its value at zero temperature, $J_z$. 
This occurs at fields above
the decoupling scale
$f\gamma_{D}^{\prime 2} = e^{1/\eta_{2D}}$, however,
which is  astronomically large and of order
$10^{38}$ at temperature below 2D melting due to
the extremely small bound on  the correlation exponent there,$^{19}$ 
$\eta_{2D} <     (28\pi)^{-1}$. 
At  large anisotropy,
$\gamma^{\prime} > \gamma_D^{\prime}$,
the system is best described by an entangled
stack of 2D vortex lattices$^{11}$ that exhibit
a relatively small macroscopic Josephson effect.$^{8}$ 
Last, the CG ensemble (2) indicates that a 3D
vortex-lattice melting transition
occurs at an intermediate temperature $T_m$ when the typical distance between
neighboring dual   charges, $n_z = \pm 1$, grows to be of order
$\xi_{2D}$, at which point  
these charges are confined into neutral pairs.$^{8, 14}$
It can be shown that 
$T_m$ lies inside of the 2D-3D cross-over window
$[T_m^{(2D)},   T_{\times}]$ 
  by virtue of this definition (see ref. 8, Eq. 62). 
Also, by comparison  with the layered CG ensemble
(2) in zero field,$^{8, 19}$
the author has argued that  in the weak-coupling limit,
$T_m$ marks the location of a second-order 
melting transition that  separates the  superconducting and normal phases.
This means that $\xi_{\perp} (T)$  diverges as $T$ cools down to $T_m$.
A second-order transition in the vortex-liquid phase of YBCO 
that resembles the above has been reported recently.$^1$

Let us now  determine what happens as interlayer coupling increases from the 
weak-coupling limit just studied.
The $n_z$ charges
are screened at low temperature, $T < T_m^{(2D)}$, which means that
no phase transition can take place as a function of the
anisotropy parameter,$^{8}$  $\gamma^{\prime}$.
Instead, a
cross-over region exists for anisotropy parameters below
$\gamma_{D}^{\prime}$
that  separates a set of weakly coupled 2D vortex lattices 
at high field from a conventional 3D vortex lattice at low field.
Again, the extremely small bound
on the 2D correlation exponent $\eta_{2D}$ at temperatures
below 2D melting indicates that the former weakly coupled
phase is not attainable there in practice.
Eqs. (13) and (14) also imply that 
the Josephson effect is essentially  independent of field/anisotropy
at these temperatures, $T < T_m^{(2D)}$.  
This observation  is consistent with Monte Carlo simulation results
of the layered $XY$ model with uniform frustration.$^7$
On the disordered side, $T > T_m^{(2D)}$,
the phase correlation length across layers,
$\xi_{\perp}$ ,
 begins to grow larger than the spacing between adjacent layers
at fields below          $f\gamma_{\times}^{\prime 2}$.$^{8}$  
Outside of the 2D critical region, 
at $\xi_{2D}\sim a_{\rm vx}$, 
Monte Carlo simulations of the layered $XY$ model with
uniform frustration indicate that  first-order melting   
occurs along the  decoupling contour 
$\langle {\rm cos}\, \phi_{l,l+1} \rangle   \sim  1/2$.$^{7, 15}$    
The  resulting phase diagram
is depicted by Fig. 1.


\bigskip
\centerline
{\bf  IV. Sublimated Decoupling}
\bigskip

We shall next apply the partial duality analysis outlined in
section II to the layered $XY$ model with only moderately
small frustration.
Let us consider again an isolated $XY$ model over the square lattice, 
but with a uniform vorticity (frustration) between
$1/30  <   f  <   1/2$.  Monte Carlo simulations
indicate that a depinning transition
at $k_B T_p^{(2D)} = 1.5  f J$
now separates a 
pinned triangular vortex lattice at low-temperature
from a vortex liquid phase at high temperature.$^{16}$
The depinning transition is first order
and no signs of a ``floating'' vortex-lattice phase 
are observed.  
Strict long-range phase correlations then exist at
low temperatures $T < T_p^{(2D)}$ in the pinned phase,
which    implies  that the
phase auto-correlation functions
are given asymptotically by  Eq. (9) with $\eta_{2D} = 0$.  
Also, the disordered phase at high temperature $T > T_p^{(2D)}$  should be
hexatic due to the underlying square-lattice grid.$^{17}$
This means that the phase autocorrelations (8) exhibit  exponential decay
as well as a trivial phase factor:$^{19}$ 
$\xi_{2D} < \infty$ and
$\int_1^2  \vec A^{\prime}\cdot d\vec r = \phi_0 (2) - \phi_0 (1)$.
Below, we shall use these facts    to map out the phase
diagram of the layered $XY$ model at   such relatively high
vortex density.

The first-order nature of the depinning transition in
an isolated $XY$ layer 
with relatively large uniform  vorticity,  $1/2 > f > 1/30$,
 implies that
the phase correlation length is finite at
temperatures just above the depinning transition:
 $\xi_{2D} (T_p^{(2D)} +) < \infty$.  By Eq. (16),
the 2D-3D cross-over field here must also then be finite.  
Notice that  $f\gamma_{\times}^{\prime 2}$
is larger than unity at depinning if  $\xi_{2D} > a_{\rm vx}$
and if $g_0 \sim 1$, since $J > k_B T_p^{(2D)}$ for
$f < 1/2$.
Strict long-range phase coherence
($\eta_{2D} = 0$) exists on the low-temperature side
at  $T < T_p^{(2D)}$, however.
We therefore reach the remarkable conclusion that at large anisotropy 
parameters of the corresponding layered $XY$ model, 
$\gamma^{\prime}\gg\gamma_{\times}^{\prime} [T_p^{(2D)}]$,
the line $T = T_p^{(2D)}$
marks a sublimation transition that separates a decoupled
vortex liquid at $T > T_p^{(2D)}$ 
with essentially {\it no} interlayer phase
coherence, $\xi_{\perp} < d$,
from a pinned 3D vortex
lattice state at $T < T_p^{(2D)}$ with long-range interlayer phase
coherence, $\xi_{\perp} = \infty$.
As depicted by Fig. 2, no 2D-3D cross-over regime exists in
such case.
Also, comparison of Eqs. (13) and (14) with   the fact
that the 2D correlation exponent $\eta_{2D}$ vanishes 
in the low-temperature phase implies that the cross-over  
 at $\gamma^{\prime} = \gamma_D^{\prime} (T)$ 
between weakly coupled and moderately coupled vortex lattices
must collapse onto the depinning line at
$T = T_p^{(2D)}$ and $\gamma^{\prime} > \gamma_{\times}^{\prime}[T_p^{(2D)}]$.  
Indeed, Eq. (13) indicates that
the Josephson coupling $\langle {\rm cos}\, \phi_{l, l+1}\rangle$
is  independent of field, $f\gamma^{\prime 2}$, at temperatures
 below the sublimation transition and
at such large anisotropy parameters.
Last, the local Josephson coupling
jumps down to a small value given by the vortex-liquid result,$^{15}$
Eq. (6) at $n = 1$,
once the vortex lattice   sublimates.
Similar jumps of order unity have been observed at vortex-lattice
melting in BSCCO.$^5$
In conclusion, the three possible
decoupling transitions 
collapse onto
a single sublimation transition!  
Such point-like
as opposed to line-like melting of the vortex lattice has been
observed in   Monte Carlo simulations of the layered $XY$ model with
moderately small frustration.$^{7}$


\bigskip
\centerline
{\bf V. Discussion and Conclusions}
\bigskip

Among the important theoretical results listed above is the local
Josephson coupling in the vortex-lattice phase, Eq.  (13),
which can be expressed as
$\langle {\rm cos}\,\phi_{l,l+1}\rangle  =    
y_0 + g_0 e^{-{1\over 2} T/T_{D} (B_{\perp})}$, with a temperature scale
$k_B T_{D} (B_{\perp}) = 2\pi\bar J/{\rm ln}(B_{\perp}/B_{\perp}^*)$.
Here, $B_{\perp}^* = \Phi_0/\Lambda_0^2$ is the naive decoupling
field$^3$ and $\bar J = k_B T/2\pi \eta_{2D}$  
is the 2D phase rigidity.$^{18}$
As observed previously, the weak logarithmic
field dependence above implies that 
$\langle {\rm cos}\,\phi_{l,l+1}\rangle$
is of order unity at low temperatures $T < T_m^{(2D)}$ 
and at perpendicular  fields below the astronomically large
scale $H_D\sim 10^{38} B_{\perp}^*$.
The local Josephson coupling (13) shows essentially no field
dependence in  such case.  This is confirmed directly  by Monte Carlo
simulations of the layered $XY$ with low uniform frustration.$^{7}$
Despite the fact that the decoupled  vortex-lattice state
characterized by a small ``cosine''
does not exist     in practice at temperatures below 2D melting,
it is nevertheless remarkable that $T_{D}(B_{\perp})$ coincides,
to within a large numerical constant,
with the temperature scale for layer decoupling induced by
the unbinding of topological
defects of the vortex lattice known as ``quartets''. 
These consist of two opposing  dislocation pairs in parallel
inside of a given layer.$^{2, 11}$ 
Comparison with the present results then
 indicates that layer decoupling is indeed
due to   such a ``quartet'' unbinding mechanism, 
but that this occurs only for exponentially weak Josephson coupling 
at temperatures below 2D melting (cf. ref. 12).
Glazman and Koshelev have  also  calculated 
the local Josephson coupling 
$\langle {\rm cos}\,\phi_{l,l+1}\rangle$
within the
3D elastic medium description for the vortex lattice,$^3$
where they find  
a much stronger dependence
$T_{D}^{\prime} (B_{\perp}) \sim (B_{\perp}^*/B_{\perp})^{1/2} T_m^{(2D)}$ 
for the  decoupling temperature scale with field, 
on  the other hand.$^{2}$
This discrepancy is due to the fact that 
the  elastic-medium approximation
 represents a continuum theory.
It therefore accounts only for long-wavelength fluctuations of the phase
difference across layers.
In the weak-coupling limit, the
dominant contribution to the ``cosine'' 
is due to short-wavelength phase fluctuations 
between adjacent layers.
These fluctuations  are missed
by the 3D elastic medium approximation,
and we believe that this is why the Glazman-Koshelev
result$^3$ underestimates the
size of the decoupling temperature scale at weak coupling.

In conclusion, a partial duality analysis of the layered $XY$ model
with uniform frustration finds that sublimated melting/decoupling
of the 3D vortex lattice occurs if
({\it i}) the superconducting-normal transition of an isolated
layer is first-order 
and if ({\it ii}) a sufficient degree 
of layer anisotropy exists.
Condition  ({\it i}) is gauranteed at strong 
substrate pinning,$^{16}$
 $1/2 > f > 1/30$.
It has also been emphasized that no decoupled  vortex-lattice state       
exists at temperatures below 2D ordering 
except for exponentially weak Josephson coupling between layers
(see Figs. 1 and 2).
This is notably consistent with complementary calculations 
that include interlayer magnetic coupling, but that
turn off the Josephson coupling.$^{12}$
It must be mentioned, however, that the magnetic coupling
between layers is weak in the extreme type-II regime studied here,
and that this coupling
can in fact be incorporated into the present duality
analysis (2) of the vortex lattice in layered superconductors
via an effective ``substrate potential'' for    isolated layers
(see ref. 12).
The   additional substrate consists of an array of commensurate pins
that mimics      the
magnetic effect of the vortex lattice in adjacent layers. 
It can therefore only 
increase phase coherence (3) inside of each 2D vortex lattice.$^{20}$
This means that the bound, $\eta_{2D} < (28\pi)^{-1}$,
on the phase  correlation exponent of the 2D vortex lattice
continues to hold. 
Hence, within the ``substrate potential'' approximation
for magnetic coupling,$^{12}$
the decoupling crossover to an entangled vortex lattice$^{11}$
with $\rho_s^{\perp}\ll J_z$
does not occur in practice
at temperatures below 2D melting
in the extreme type-II regime [see Eq. (14)].
We remind the reader that rigid translations of the vortex
lattice are assumed throughout to be prohibited by surface barriers
(see  ref. 19).

The author is grateful for the hospitality of the
Instituto de Ciencias de Materiales de Madrid, where
this work was completed, and to Marty Maley
and Paco Guinea for discussions.

\vfill\eject
\centerline{\bf References}
\vskip 16 pt



\item {1.} 
F. Bouquet, C. Marcenat, E. Steep, R. Calemczuk, W.K. Kwok,
U. Welp, G.W. Crabtree, R.A. Fisher, N.E. Phillips and A. Schilling,
Nature {\bf 411}, 448 (2001).

\item {2.}  G. Blatter, M.V. Feigel'man, V.B. Geshkenbein, A.I. Larkin,
and V.M. Vinokur, Rev. Mod. Phys. {\bf 66}, 1125 (1994).

\item {3.} L.I. Glazman and A.E. Koshelev, Phys. Rev. B {\bf 43}, 2835 (1991).

\item {4.} D.T. Fuchs, R.A. Doyle, E. Zeldov, D. Majer, W.S. Seow,
 R.J. Drost, T. Tamegai, S. Ooi, M. Konczykowski and P.H. Kes,
Phys. Rev. B {\bf 55}, R6156 (1997);

\item {5.}
T. Shibauchi, T. Nakano, M. Sato, T. Kisu, N. Kameda, N. Okuda, S. Ooi
and T. Tamegai,
Phys. Rev. Lett. {\bf 83}, 1010 (1999);
M.B. Gaifullin, Y. Matsuda, N. Chikumoto, J. Shimoyama and K. Kishio,
Phys. Rev. Lett. {\bf 84}, 2945 (2000).

\item {6.} E.F. Righi, S.A. Grigera, G. Nieva, D. Lopez and F. de la Cruz,
Phys. Rev. B {\bf 55}, 14156 (1997);
X.G. Qiu, V.V. Moshchalkov and J. Karpinski, 
Phys. Rev. B {\bf 62}, 4119 (2000).

\item {7.} A.E. Koshelev, Phys. Rev. B {\bf 56}, 11201 (1997).

\item {8.} J.P. Rodriguez, Phys. Rev. B {\bf 62}, 9117 (2000);
Physica C {\bf 332}, 343 (2000); Europhys. Lett. {\bf 54}, 793 (2001).

\item {9.} S. Hikami and T. Tsuneto, Prog. Theor. Phys. {\bf 63}, 387 (1980);
C. Kawabata, M. Takeuchi, S.R. Shenoy and A.R. Bishop, J. Phys. Soc. Jpn.
{\bf 69}, 194 (2000).

\item {10.} L.I. Glazman and A.E. Koshelev, Zh. Eksp. Teor. Fiz.
 {\bf 97}, 1371 (1990) [Sov. Phys. JETP {\bf 70}, 774 (1990)].

\item {11.} M. Feigel'man, V.B. Geshkenbein, and A.I. Larkin, 
Physica C {\bf 167}, 177 (1990); 
E. Frey, D.R. Nelson, and D.S. Fisher, 
Phys. Rev. B {\bf 49}, 9723 (1994).


\item {12.} M.J.W. Dodgson, V.B. Geshkenbein and G. Blatter,
Phys. Rev. Lett. {\bf 83}, 5358 (1999).

\item {13.}  S.E. Korshunov, Europhys. Lett. {\bf 11}, 757 (1990).

\item {14.} A. Polyakov, Phys. Lett. {\bf 72} B, 477 (1978).

\item {15.} A.E. Koshelev, Phys. Rev. Lett. {\bf 77}, 3901 (1996).


\item {16.} M. Franz and S. Teitel, Phys. Rev. B {\bf 51}, 6551 (1995);
S.A. Hattel and J.M. Wheatley, Phys. Rev. B {\bf 51}, 11951 (1995).


\item {17.} D.R. Nelson and B.I. Halperin, Phys. Rev. B {\bf 19},
2457 (1979).

\item {18.} J.M. Kosterlitz, J. Phys. C {\bf 7}, 1046 (1974).

\item {19.} J.P. Rodriguez, Phys. Rev. Lett. {\bf 87}, 207001 (2001).

\item {20.} C.E. Creffield and J.P. Rodriguez,
``Optimum Pinning of the Vortex Lattice in Extremely Type-II Layered
Superconductors'' (cond-mat/0205231).





\vfill\eject
\centerline{\bf Figure Captions}
\vskip 20pt
\item {Fig. 1.}   Shown is the proposed phase diagram for the layered
$XY$ model with uniform frustration in the continuum regime, $f\ll 1/36$.
Notice the absence (in practice) of a 	decoupling transition at
temperatures below 2D melting.
Rigid translations of the vortex lattice are assumed to be prohibited by
surface barriers.  The mean-field temperature dependence
$J\propto T_{c0} - T$ is also assumed.

\item {Fig. 2.} The proposed phase diagram for the layered $XY$ model
with moderate uniform frustration, $1/30 < f < 1/2$ is displayed.
The mean-field temperature dependence
$J\propto T_{c0} - T$ is assumed once again.

\vfill\eject
%
\magnification=1200
\baselineskip=17pt

\centerline {\bf Erratum:   ``Sublimated decoupling of the vortex lattice}
\centerline
{\bf \quad\qquad\qquad in extremely type-II layered superconductors'',}
\centerline
{\bf \quad\qquad[Phys. Rev. B {\bf 66}, 214506 (2002)]}
\centerline
{\it \quad\qquad  J.P. Rodriguez}
\bigskip\bigskip\bigskip
The decoupling field
for temperatures  that lie below the 2D  ordering transition
that was derived in the discussion following Eq.  (16)
is more generally given by
$$f\gamma_D^{\prime 2} = (r_0 / a_{\rm vx})^2 e^{1/\eta_{2D}},$$
where $r_0\sim a_{\rm vx}$ was implicitly assumed.
Although the latter is not necessarily true,
the ratio $r_0 / a_{\rm vx}$ must be larger than
$\kappa^{-1} = \xi_0 /\lambda_0$.
We have $\kappa\sim 100$ in YBCO for example.
The above then implies that
the decoupling field is bounded by
$f\gamma_D^{\prime 2} > 10^{34}$ at temperatures below 2D ordering
in such case, since $\eta_{2D} < (28\pi)^{-1}$.
It therefore remains exponentially big.

More seriously, the claim made in section IV that the
2D  phase  correlation exponent is null at temperatures that
lie below the 2D vortex-lattice depinning transition
is incorrect.
What is null  is its vortex component,
which leaves  the spin-wave result
$\eta_{2D} = k_B T / 2\pi J$ for the net exponent.
The sentences in the middle of both paragraphs of section IV
that begin with
``Strict long-range phase ...''
must therefore be replaced with
``Quasi long-range phase ...''.
Also, the equation ``$\eta_{2D} = 0$''
that appears in
both of these sentences
must be replaced with ``$\eta_{2D} = k_B T / 2\pi J$''.
The rest of section IV remains valid for Josephson
coupling that is not exponentially weak.
The equation displayed above,
for example,
yields an astronomically large  lower bound
$f\gamma_D^{\prime 2} > (r_0 / a_{\rm vx})^2 \cdot 10^{45}$
for the decoupling field
at temperatures below 2D ordering
and at an in-plane vortex concentration of $f = 1 / 25$.
This bound is due to the value
$k_B T_p^{(2D)} = 0.06 J$
of the  first-order  transition temperature of an isolated layer
in such case.

The above corrections do not change any of the
conclusions drawn in the paper.

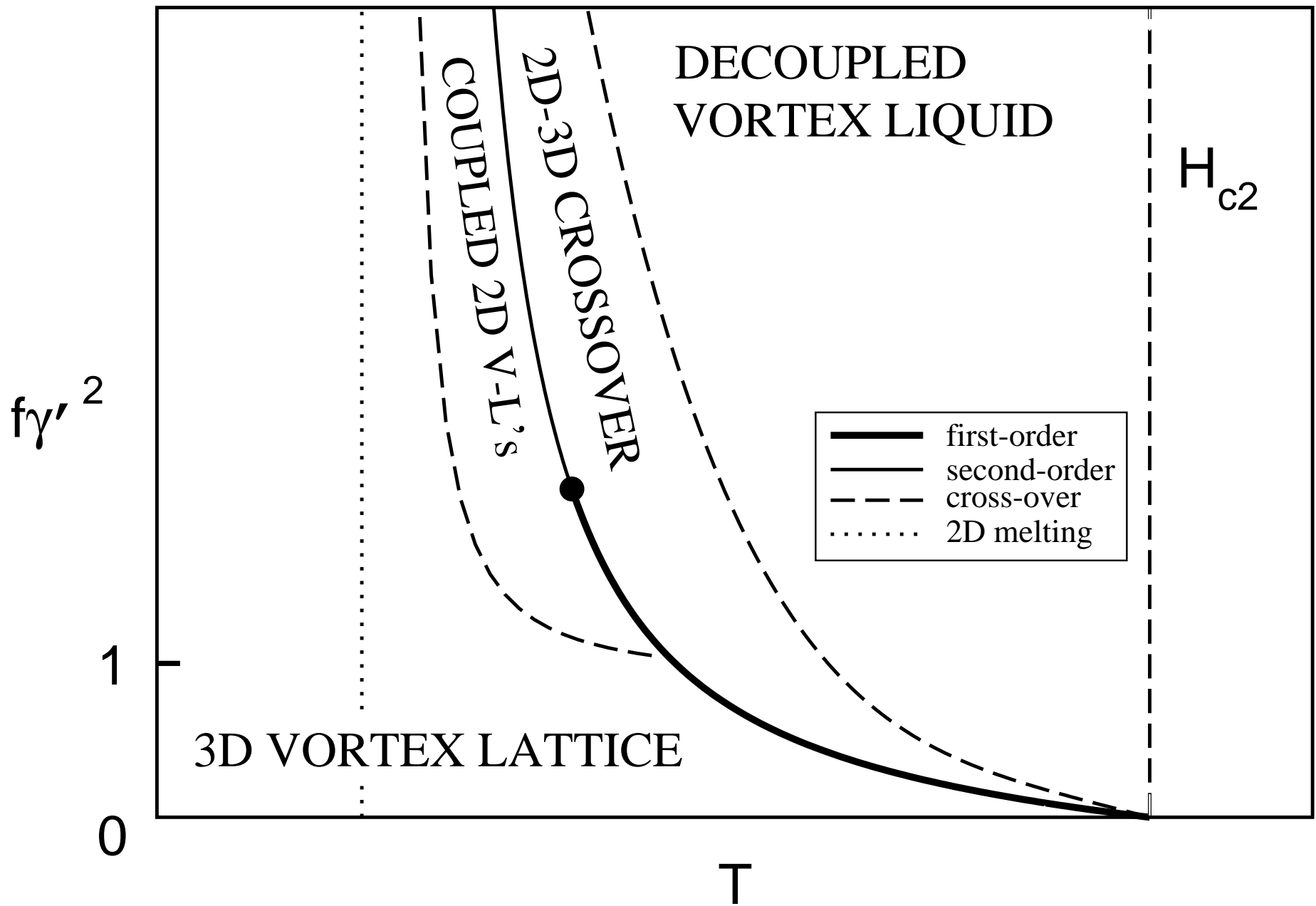
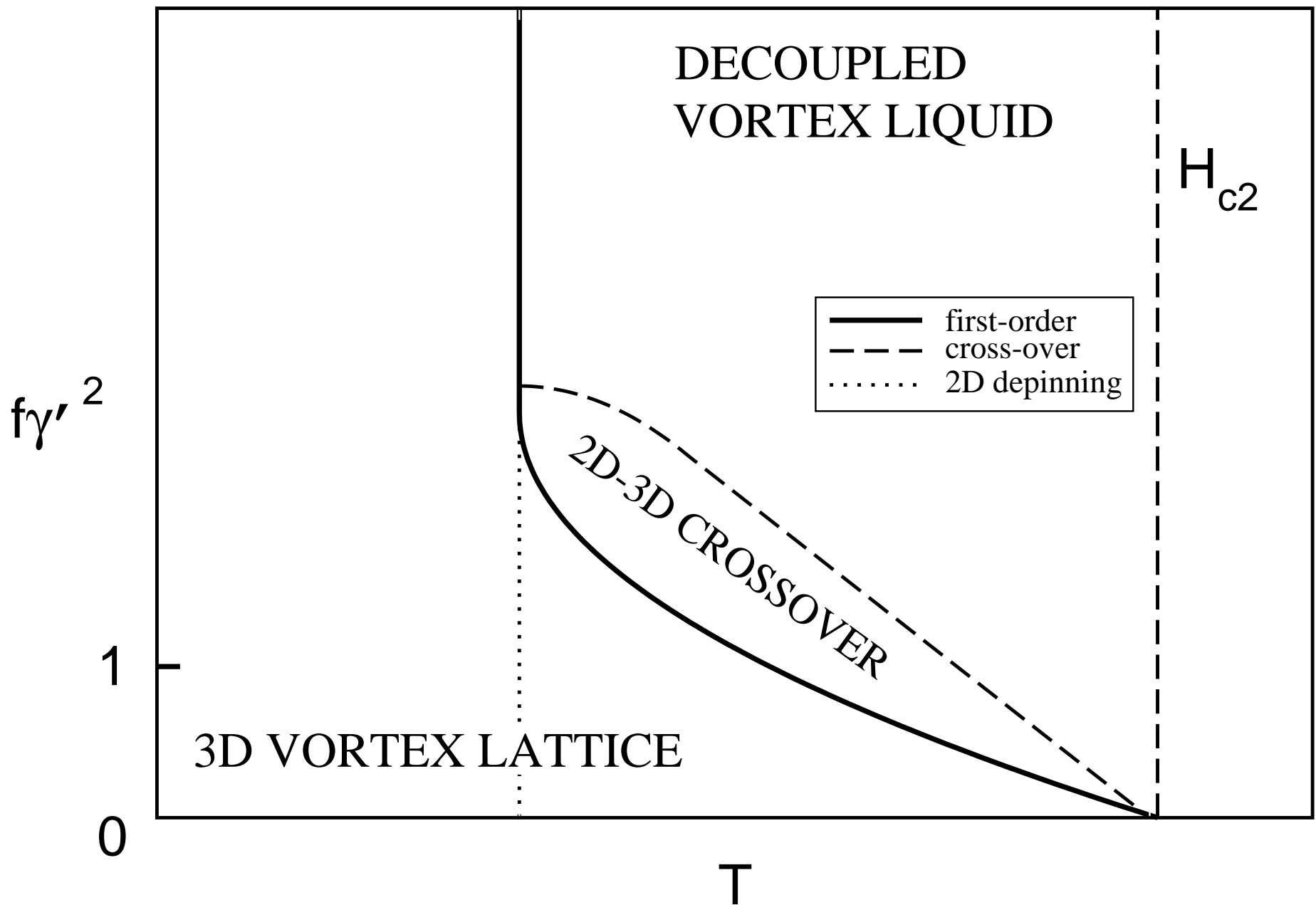
\end